%% file: main.tex
\documentclass[12pt]{iopart}

\usepackage{graphicx}
\usepackage{float}
\usepackage{hyperref}
\usepackage{tikz}
\usetikzlibrary{svg.path}
\usepackage{orcidlink}
\hypersetup{
    colorlinks=true, 
    pdfborder = {0 0 0.5 [3 3]},
    anchorcolor=black,
    citecolor=blue,
    linktoc=all,    
    linktocpage=true,
    linkcolor=red,
	urlcolor=blue
}

\usepackage{color}   
\usepackage{xcolor}
\begin{document}

\title[A directed CW search from NS in binary systems]{A directed continuous-wave search from Scorpius X-1 with the five-vector resampling technique}

\author{Francesco Amicucci\orcidlink{0009-0005-2139-4197}$^{1,2}$, Paola Leaci\orcidlink{0000-0002-3997-5046}$^{1,2}$, Pia Astone\orcidlink{0000-0003-4981-4120}$^{1}$, Sabrina D'Antonio\orcidlink{0000-0003-0898-6030}$^{^3}$, Stefano Dal Pra$^4$\orcidlink{0000-0002-1057-2307}, Matteo Di Giovanni$^{1,2}$\orcidlink{0000-0003-4049-8336}, Luca D'Onofrio\orcidlink{0000-0001-9546-5959}$^{1}$, Federico Muciaccia$^{1,2}$\orcidlink{0000-0003-0850-2649}, Cristiano Palomba\orcidlink{0000-0002-4450-9883}$^1$, Lorenzo Pierini$^{1}$\orcidlink{0000-0003-0945-2196}, Akshat Singhal$^{5}$\orcidlink{0000-0003-1275-1904}}
\address{$^1$INFN, Sezione di Roma, I-00185 Roma, Italy
\\$^2$Università di Roma “La Sapienza", I-00185 Roma, Italy
\\$^3$INFN, Sezione di Roma Tor Vergata, 00133 Roma, Italy
\\$^4$INFN, CNAF, 40127 Bologna, Italy
\\$^5$HBCSE, Tata Institute of Fundamental Research, Mumbai 400088, India}

\ead{francesco.amicucci@roma1.infn.it and paola.leaci@roma1.infn.it}

\vspace{10pt}
\begin{indented}
\item[]March 2024
\end{indented}

\begin{abstract}
Continuous gravitational-wave signals (CWs), which are typically emitted by rapidly rotating neutron stars with non-axisymmetric deformations, represent particularly intriguing targets for the Advanced LIGO-Virgo-KAGRA detectors. These detectors operate within sensitivity bands that encompass more than half of the known pulsars in our galaxy existing in binary systems (i.e., over 417 pulsars), which are the targeted sources of this paper. However, the detection of these faint signals is especially challenged by the Doppler modulation due to the source's orbital motion, typically described by five Keplerian parameters, which must be determined with high precision to effectively detect the signal. This modulation spreads the signal across multiple frequency bins, resulting in a notable reduction of signal-to-noise ratio and potentially hindering signal detection. To overcome this issue, a robust five-vector resampling data-analysis algorithm has been developed to conduct thorough directed/narrowband CW searches at an affordable computational cost. We employ this methodology for the first time to search for CWs from Scorpius X-1, using publicly available data from the third observing run of the Advanced LIGO-Virgo-KAGRA detectors. No statistically significant CW signals can be claimed. Hence, we proceeded setting 95\% confidence-level upper limits in selected frequency bands and orbital parameter ranges, while also evaluating overall sensitivity.
\end{abstract}

\input{1_introduction}
\input{2_Signal_model}
\input{3_Resampling}
\input{4_Results}
\input{5_sensitivity}
\input{6_conclusions}
\input{7_Acknowledgments}
\clearpage

\appendix

\input{A_appendix}
\input{B_appendix}
\input{C_appendix}
\clearpage
\phantomsection

\section*{References}
\bibliographystyle{iopart-num}
\bibliography{main}
\end{document}

%% file: 1_introduction.tex
\section{Introduction}
Gravitational waves (GWs) are a fundamental prediction of Albert Einstein's General Theory of Relativity \cite{1916SPAW.......688E, 1918SPAW.......154E} and are emerging as a revolutionary breakthrough in astrophysics. These ripples in the fabric of spacetime carry crucial information on the dynamics and interactions of massive celestial objects, providing a new lens through which we can explore the universe. Since their first direct detection in 2015 \cite{PhysRevLett.116.061102} by the Laser Interferometer Gravitational-Wave Observatory (LIGO) \cite{Aasi_2015} and the Virgo collaboration \cite{Acernese_2015}, GWs have revolutionized our understanding of the cosmos, offering unprecedented insights into the most energetic and cataclysmic events in the universe \cite{Abbott_2019}.

Understanding GWs not only advances our knowledge of astrophysics, but also presents opportunities to test and refine our understanding of the Theory of Gravity.

Unlike transient GWs emitted by the merger of compact objects, there are much fainter, persistent/continuous GWs (CWs), which are being continuously emitted by a source that has quasi-periodic rotational frequency. Due to their weakness, we need to integrate for long observing times to accumulate an adequate signal-to-noise-ratio (SNR) \cite{universe5110217}. Promising sources of CWs include neutron stars (NSs), the remnants of massive stellar cores after a supernova explosion, that can emit a continuous stream of GWs due to their asymmetrical shapes, non-uniform rotation, or other structural irregularities \cite{PhysRevD.66.084025}. The most commonly accepted model suggests \textit{mountains} supported by elastic and/or magnetic stresses, where the NS rotational frequency $f_{\rm rot}$ is related to the GW frequency $f_{\rm GW}$ by $f_{\rm GW}=2f_{\rm rot}$ \cite{PhysRevD.20.351}. Detecting CWs would allow us to get a set of information that electromagnetic observations alone cannot achieve, such as information on NS quadrupolar deformation (i.e., ellipticity), NS properties (i.e., the range of NS masses, radii, population models), and in general cold dense matter Equation Of State properties \cite{universe5110217}. A review of the last twenty years of efforts to detect CWs can be found in \cite{WETTE2023102880}.

In this paper, we focus on Scorpius X-1, which is the brightest Low-Mass X-ray binary (LMXB) composed of a NS in a binary orbit with a low-mass normal star (i.e., its companion). A characteristic property of these systems is the inflow of gas from the companion star to the NS, in a process known as accretion \cite{10.1093/mnras/184.3.501, Bildsten_1998}. The by-product of accretion is the generation of X-rays: the more accretion, the more X-rays are produced.

An interesting astrophysical model predicts that the GW strain $h_0$ from a LMXB at the torque-balance level is proportional to the square root of the x-ray flux $F_x$ \cite{1984ApJ_278_345W, Bildsten_1998, 10.1093/mnras/184.3.501}. Since Scorpius X-1 is the brightest LMXB known so far, we consider it as one of the most interesting potential CW sources.
Scorpius X-1 ephemerides (obtained through electromagnetic observations) are summarized in \tref{tab:ScoX1-ephe}. 

\begin{table}[H]
\caption{\footnotesize\label{tab:ScoX1-ephe}Observed Parameters of the LMXB Scorpiux X-1.}
\begin{indented}
\lineup
\item[]
\begin{tabular}{@{}*{5}{l}}
\br  
         Parameter & & Value& Units&Ref  \cr
         \mr
         right ascension$^{\rm a}$ & $\alpha$ & 16h 19m 55.067s  & & \cite{Singhal_2019}\cr
         declination & $\delta$ & $-15^\circ$ $38^{\prime} 25.02^{\prime\prime}$    &  & \cite{Singhal_2019} \cr
         frequency & $f_0$ & (unknown) & Hz & \cr
         projected semi-major axis & $a_p$    &  [1.45, 3.25] & ls  & \cite{10.1093/mnras/sty1441}\cr
         time of ascending node & $t_{\rm asc}$ & 1078153676 $\pm$ 33 & GPS s & \cite{10.1093/mnras/stad366} \cr
         & & 06/03/2014 15:07:40 & UTC\cr
         orbital period$^b$ & $P$ & 68023.92 $\pm$ 0.02 & s & \cite{10.1093/mnras/stad366} \cr
         & & 0.7873139 $\pm$ 0.0000007 & d\cr
         eccentricity & $e$ & $\leq 0.0132$ & & \cite{10.1093/mnras/stad366}\cr
         argument of periapsis & $\omega$ & $[0,2\pi]$ (unknown) & rad & \cr
\br
    \end{tabular}
    \item \textbf{Notes.} Uncertainties are $1\sigma$ unless otherwise stated.
    \item[] $^{\rm a}$ The sky position is determined to the microarcosecond and therefore can be treated as known in the present search.
    
\end{indented}
\end{table}

Several CW searches targeting Scorpius X-1 have been conducted so far, such as the Viterbi “hidden Markov Model” method \cite{PhysRevD.106.062002, vargas2023searchgravitationalwavesscorpius}, and the cross-correlation CrossCorr search \cite{Abbott_2022_scox1, Whelan_2023}, which was recently improved by using efficient lattice-based template banks \cite{Wagner_2022} and resampling techniques for efficient frequency computations \cite{PhysRevD.97.044017, Zhang_2021}. More recently, the resampling has been also used to develop a new search pipeline called \texttt{BINARYWEAVE} that employs a semicoherent $\mathcal{F}$-statistic StackSlide approach \cite{PhysRevD.107.062005}.
In the current work, we perform a directed/narrowband\footnote{The search is narrowband in the sense that it can account for nonzero uncertainties in the orbital Keplerian parameters \cite{Singhal_2019}} search for CWs from Scorpius X-1, using the binary five-vector methodology \cite{Singhal_2019} to analyse, for the first time with this method, data from the third observing LIGO-Virgo-KAGRA science run (O3)\footnote{O3 started on April 1, 2019 15:00:00 UTC (GPS 1238166018) and ended on March 27, 2020 17:00:00 UTC (GPS 1269363618).}
\cite{Abbott_2023,data_Sets}.

The paper is organized as follows. In section \ref{signal-model} we summarize the main features of CWs and how they will be observed at the detector site. In section \ref{resa} we describe the five-vector resampling method for NSs in binary systems. In section \ref{results} we present the results for a narrowband CW search from Scorpius X-1 using the O3 data. In section \ref{sec:sensitivity} we show  95\% confidence-level upper limits in selected frequency bands and orbital parameter ranges. Finally, we summarize the main outcomes in section \ref{sec:conclusions}.

%% file: 2_Signal_model.tex
\section{Signal model}\label{signal-model}
The expected GW signal $h(t)$ from a nonaxisymmetric NS, at the time of arrival at the detector $t=t_{\rm arr}$, can be written using the polarization ellipse as the real part of
\cite{Astone_2010}
\begin{eqnarray}\label{eq:strain-with-H0}
h(t_{\rm arr})=H_0\left(H_+A_+(t_{\rm arr})+H_\times A_\times(t_{\rm arr})\right)e^{{\rm i} \Phi_{\rm det}(t_{\rm arr})}\,,
\end{eqnarray}
where $\Phi_{\rm det}$ is the signal phase in the detector frame, $H_0$ is the wave amplitude, $A_{+,\times}(t)$ are the time-dependent sidereal detector response functions and are linked to the antenna patterns functions; $H_{+,\times}$ are the complex polarization functions, which are given by
\begin{eqnarray}\label{eq:complex-pol}
    H_+&=\frac{\cos(2\psi)-{\rm i}\eta_{\rm p}\sin(2\psi)}{\sqrt{1+\eta^2_{\rm p}}},\\
    H_\times&=\frac{\sin(2\psi)+{\rm i}\eta_{\rm p}\cos(2\psi)}{\sqrt{1+\eta^2_{\rm p}}},
\end{eqnarray}
where $\psi$ is the polarization angle, defined as the orientation of the major axis of the polarization ellipse relative to the celestial parallel of the source, measured counterclockwise, and $\eta_{\rm p}$ corresponds to the CW degree of polarization and is defined as
\begin{eqnarray}
    \eta_{\rm p}=-\frac{2\cos \iota_{\rm rot}}{1+\cos^2(\iota_{\rm rot})}\,,
\end{eqnarray} 
with $\iota_{\rm rot}$ being the inclination angle of the star's rotation axis with respect to the line of sight. The $\eta_{\rm p}$ parameter varies in the range $[-1, 1]$, where $\eta_{\rm p} = 0$ for a linearly polarized wave and $\eta_{\rm p} = \pm 1$ for a circularly polarized wave ($\eta_{\rm p} = 1$ if the circular rotation is counter-clockwise). 

The two complex amplitudes satisfy the condition $|H_+|^2 + |H_\times|^2 = 1$. The time dependent sidereal detector response functions $A_+(t),A_\times (t)$ depend on the relative position between the source and the detector, which changes due to the Earth sidereal motion. Hence, they encode the signal amplitude modulation. For a ground based \textit{L-shaped} detector, the sidereal response can be precisely characterized by five components in the Fourier transform, with the Fourier coefficients determined by the source position, as well as the detector location and orientation on Earth \cite{D'Antonio_2009}. As discussed in \cite{Astone_2010}~\eref{eq:strain-with-H0} is equivalent to the standard expression given by (4) in \cite{PhysRevD.58.063001}. As it can be noticed, in~\eref{eq:strain-with-H0} we work with $H_0$ rather than the classic strain amplitude $h_0$, which -for the ideal case of a steadily spinning triaxial ellipsoid star emitting GWs only at twice the rotation frequency- is given by (see, e.g.,~\cite{Abbott_2022}):
\begin{eqnarray}\label{eq:norm-ho}
h_0&= \frac{16\pi^2 G}{c^4rP^2}I_{\rm zz}\epsilon=\\
&=1.05\times 10^{-27}\left[\frac{f_{{\rm GW}}}{{100\,\rm Hz}}\right]^2\left[\frac{10{\rm  kpc}}{r}\right]\left[\frac{I_{\rm zz}}{10^{38}\,{\rm  kg\,m}^2}\right]\left[\frac{\epsilon}{10^{-6}}\right]\,.
\end{eqnarray}
We note that $r$is the distance to the source, $I_{zz}$ is the source's moment of inertia along the star rotation axis $\hat{z}$, 
$c$ is the speed of light, $G$ is the universal gravitational constant and $\epsilon$ is the ellipticity, which is a measure of the NS deformation, defined as
\begin{eqnarray}
    \epsilon\equiv \frac{|I_{\rm xx} - I_{\rm yy}|}{I_{\rm zz}}.
\end{eqnarray}
The relation between these two amplitudes is the following \cite{Astone_2010}
\begin{eqnarray}\label{eq:Ho-h0}
H_0=\frac{h_0}{2}\sqrt{1+6\cos^2\iota_{\rm rot}+\cos^4\iota_{\rm rot}}\,,
\end{eqnarray}
which is equal to $h_0/2$ for linear polarization and $h_0\sqrt{2}$ for circular polarization. 
We prefer using $H_0$ to avoid disentangling the information between $h_0$ and $\iota_{\rm rot}$. As shown in (\ref{eq:strain-with-H0}), $h(t)$ is the product of two terms: the first, $\left(H_+A_+(t)+H_\times A_\times(t)\right)$, is a ‘slow’ amplitude modulation, while the second, $e^{{\rm i}\Phi(t)}$, is a ‘fast’ term, due to the intrinsic source frequency \cite{Astone_2010}. We use the method first presented in \cite{Astone_2010} to describe both signals and data in the Fourier domain at the five frequencies $f_{\rm GW},f_{\rm GW}\pm f_{\rm sd},f_{\rm GW}\pm 2f_{\rm sd}$, where $f_{\rm sd}$ is the sidereal frequency\footnote{The Earth sidereal frequency is the inverse of sidereal period, which is $\approx86164$ s ($\approx 23 {\rm  h }\; 56 {\rm m }\; 4 {\rm s}$).} ($1.16 \times 10^{-5}$ Hz). 

%% file: 3_Resampling.tex
\section{The robust resampling technique}\label{resa}
In principle, CWs can be detected by a single GW interferometer and are persistently present into any LIGO-Virgo-KAGRA dataset, allowing for detailed analysis. The way to search for CWs depends on the amount of information available about the source. The five-vector method is a frequentist pipeline \cite{Astone_2010, D’Onofrio_2025} based on the splitting at the detector of the expected GW frequency $f_{\rm GW}$ in five components due to the sidereal modulation of the Earth. This method is robust as it enables the detection of a signal with a given SNR, even in the presence of significant uncertainties in the orbital parameters \cite{Singhal_2019}.

As discussed in \cite{Singhal_2019}, a monochromatic GW signal undergoes multiple time-dependent modulations when it reaches the ground-based detector. Our goal is to compute the time series in a new time coordinate $t'$, using the \textit{stroboscopic resampling}, i.e., the time series is downsampled at irregular intervals where the phase exhibits linear behavior. In other words, assuming that the original time series is evenly spaced, it is downsampled at the nearest integer values of the new time variable $t'$. It is important to note that, in the absence of spindown terms, stroboscopic resampling can be performed without prior knowledge of the frequency, as the time correction is independent of it. 

The (monochromatic) signal phase in the (NS) source frame is
\begin{eqnarray}
\Phi^{\rm NS}(\tau)=\phi_0+2\pi f_{\rm GW}(\tau-\tau_0)\,,
\end{eqnarray}

where $\tau$ is the source emission time and $\phi_0$ is the phase at the reference time $\tau_0$.

The signal phase observed at the detector frame is\footnote{This expression, as reported e.g. in \cite{WETTE2023102880}, is valid in the non-relativistic approximation $v/c\ll 1$, where $v$ is the detector velocity with respect to the source. Since the heliocentric velocity of the detector is $|\mathbf{v}|/c\sim 10^{-4}$ and Scorpius X-1 linear heliocentric velocity is $|\mathbf{v}_{\rm NS}|/c\sim 10^{-4}$ \cite{2020yCat.1350....0G, 1995A&AS..114..269D}, we can consider this approximation valid.
}
\begin{eqnarray}\label{eq:phase-det}
\fl
\Phi^{\rm det}(t_{\rm arr})=\phi_0+2\pi f_{\rm GW}\left(t_{\rm arr}+\frac{\mathbf{x}(t_{\rm arr})\cdot \hat{\mathbf{r}}}{c}-\frac{d}{c}-\frac{R(t_{\rm arr})}{c}+\Delta_{\rm Ein,\odot}-\Delta_{{\rm S},\odot}-\tau_0\right),
\end{eqnarray}
where $\frac{\mathbf{x}(t_{\rm arr})\cdot \hat{\mathbf{r}}}{c}$ is the so-called Earth Rømer delay term, with $\mathbf{x}(t_{\rm arr})$ being the vector from the \textit{Solar System Barycentre} (SSB) to the detector, $\hat{\mathbf{r}}$ the unit vector from the SSB towards the \textit{Binary System Barycentre} (BSB); $d/c$ is the NS delay term with $d$ the distance of the BSB to the SSB, $R(t)/c$ is the binary Rømer delay term \cite{PhysRevD.91.102003, Leaci_2017}, with $R(t)$ being the radial distance of the CW-emitting NS from the BSB projected along the line of sight; while $\Delta_{\rm Ein,\odot}$ and $\Delta_{{\rm S},\odot}$ are the Earth Einstein and Shapiro time delay, respectively \cite{PhysRevD.89.062008, 1989ApJ_345_434T}. The term $d/c$ is a constant term and can be neglected by redefining the initial phase $\phi_0$. The binary delay $R(t)$ can be expressed in terms of five Keplerian parameters $\{P,a_{\rm p}, e, \omega, t_{\rm p}\}$, where $a_{\rm p}$ is the projected semi-major axis in ls, $P$ is the time period required by a NS to complete its orbit around the BSB, $e$ is the eccentricity, $\omega$ is the argument of periapsis, and $t_{\rm p}$ is the time of periapsis passage (see \cite{PhysRevD.91.102003} and references therein). In the case of low eccentric orbit ($e\ll 1$), $R(t)$ can be equivalently parametrized by $\{\Omega, a_{\rm p}, \kappa, \eta, t_{\rm asc}\}$, where $\eta\equiv e\sin\omega$ and $\kappa\equiv e\cos\omega$ represent the \textit{Laplace-Lagrange} parameters, $t_{\rm asc}$ is the time of ascending node and $\Omega$ is the mean orbital angular velocity. Hence, we can introduce the new time variable $t'$ such that the phase $\Phi^{\rm det}(t')$ is equal to the signal with constant frequency:
\begin{eqnarray}
\Phi^{\rm det}(t')=\phi_0+2\pi f_{\rm GW}(t'-\tau_0),
\end{eqnarray}
with
\begin{eqnarray}\label{eq:new-time-var}
t' = t_{\rm arr}+\frac{\mathbf{x}(t_{\rm arr})\cdot \hat{\mathbf{r}}}{c}-\frac{d}{c}-\frac{R(t_{\rm arr})}{c}+\Delta_{{\rm Ein},\odot}-\Delta_{{\rm S},\odot}.
\end{eqnarray}
It is evident that, by resampling the data according to \eref{eq:new-time-var}, we can remove all modulations. In a hypothetical scenario, where both the detector and source are relatively stationary, i.e., all time correction terms are constant $t'(t_{\rm arr})=$ const., there will be no modulation, and therefore the signal will be monochromatic at the detector \cite{Singhal_2019}. 
As previously mentioned, the key advantage of using a rescaled time for Doppler correction in narrowband searches is that $t'$ does not depend on the frequency. This means that a single correction applies to all frequencies, making it especially useful for searching for a signal from Scorpius X-1, where the spin frequency is unknown. As explained in Section III of \cite{PhysRevD.89.062008}, for isolated sources, and in Appendix B of \cite{Singhal_2019} for binaries NSs, we can take into account the spindown correction in a similar manner, although such correction is frequency-dependent. 
In this study, we assume no spindown, based on the expected steady-state torque balance scenario in LMXBs, which are our primary focus.

\begin{figure}[H]
    \begin{indented}
    \lineup
    \item[]
    \includegraphics[width=1\linewidth]{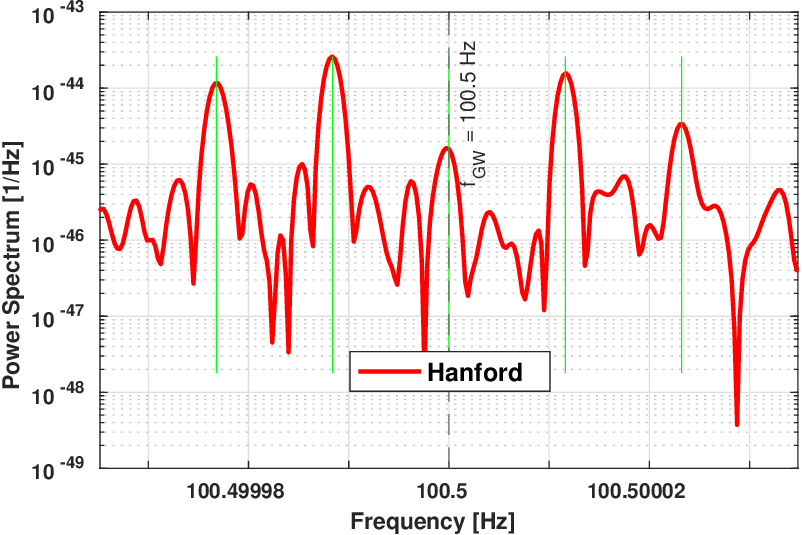}
    \end{indented}
    \caption[Power spectrum of an injected signal in the O3 Hanford to illustrate five peaks]{Power spectrum of a software injection Scorpius X-1 like signal at $100.5$ Hz and with amplitude of $1\times 10^{-24}$ for $T_{\rm obs}=10$ days. We see five peaks in the frequency domain after the time-domain correction.}
    \label{fig:5-peaks}
\end{figure}

Once that we have properly demodulated the data to correct for both the Earth Doppler effect and binary orbital motion, the GW signal power is spread among five frequencies, related to the detector sidereal responses $A_{+,\times}$ in \eref{eq:strain-with-H0}. For purely illustrative purposes, \fref{fig:5-peaks} shows the power spectrum of O3 Hanford data where a software-injected signal has been added with $H_0 = 1\times 10^{-24}$ for $T_{\rm obs}=10$ days at a frequency $100.5$ Hz, with sky location and orbital parameters of Scorpius X-1. Five-peaks are clearly visible, centered around the injection frequency, with a separation corresponding to the Earth sidereal frequency $f_{\rm sd}$.

\subsection{Matched filter and Detection statistic}
CW search methods typically employ a matched-filtering approach, where a signal model is compared to the data to compute a Detection Statistic (DS) (e.g., see \cite{SFrasca_2004, PhysRevD.58.063001}). Let us indicate the data time series
\[
x(t)=h(t)+n(t),
\]
where $h(t)$ and $n(t)$ are the GW signal and noise time series, respectively. If we apply the Doppler and relativistic corrections as described before, the signal is monochromatic apart from an amplitude and phase sidereal modulation. \Eref{eq:strain-with-H0} becomes 
\begin{eqnarray}\label{eq:strain-with-H0-const}
h(t)=H_0\left(H_+A_+(t)+H_\times A_\times(t)\right)e^{{\rm i}(2\pi f_{\rm GW}t+\phi_0)}\,,
\end{eqnarray}
where $\phi_0$ is the phase at the time $t = 0$. Given a generic time series $g(t)$, we defined a five-vector as 
\begin{eqnarray}\label{eq:five-freqs}
\mathbf{G}(f)=\left(\matrix{
    G_{-2}(f)\cr
    G_{-1}(f)\cr
    G_{0}(f)\cr
    G_{1}(f)\cr
    G_{2}(f)
}\right),
\end{eqnarray}
with
\begin{eqnarray}
G_k(f)=\int_Tg(t) \exp[-2\pi i(f-kf_{\rm sd})t]{\rm d}t, \quad {\rm with}\; k=-2,\dots,2\;
\end{eqnarray}
where $T$ is the observation time.

We therefore define the data five-vector $\mathbf{X}$ and the signal template five-vectors $\mathbf{A}_{+,\times}$ for a frequency $f$ as the Fourier components at the five frequencies where the signal
power is split. We can also represent $\mathbf{X}$ in terms of $\mathbf{A}^+$ and $\mathbf{A}^\times$, i.e., the source plus and cross five-vectors, respectively, which are the five Fourier-transform components of $A_{+,\times}(t)$. We then obtain \cite{Singhal_2019}
\[
\mathbf{X}=H_0e^{{\rm i}\phi_0}\left(H_+\mathbf{A}^++H_\times \mathbf{A}^\times\right)+\mathbf{N},
\]
where $H_0$ is given by \eref{eq:strain-with-H0} and $\mathbf{N}$ is the five-vector of noise alone, which -in absence of a signal- would be \cite{Astone_2010}
\begin{eqnarray}\label{eq:noise-5-v}
\mathbf{N}=\int_Tn(t)\exp[-2\pi i(f-kf_{\rm sd})t]{\rm d}t, \quad {\rm with}\; k=-2,\dots,2\,.
\end{eqnarray}

If the sky location is known, the source templates $(\mathbf{A}^+,\mathbf{A}^\times)$ can be used as a strong signature to match these five peaks, thereby allowing us to detect the signal. However, since the complex polarization functions ($H_+,H_\times$) depend on the generally unknown $\iota_{\rm rot}$ and $\psi$ parameters, the exact source template cannot be computed. Hence, we compute the estimator for the polarization amplitudes individually by matched filtering source five-vectors (templates) with the data five-vectors \cite{PhysRevD.89.062008}. An estimator for the polarization amplitudes, as a function of frequency, is defined as the scalar product between the data five-vector and the normalized source five-vector according to \cite{Mastrogiovanni_2017, Astone_2010}
\begin{eqnarray}\label{eq:estimators}
\hat{h}^{+,\times}(f)=\frac{\mathbf{X}(f)\cdot \mathbf{A}^{+,\times}}{|\mathbf{A}^{+,\times}|^2}\,,
\end{eqnarray}
where $\hat{h}^{+,\times}(f)$ are estimators for the $+,\times$ polarization amplitudes, respectively; the symbol ‘$\cdot$’ implies taking the complex conjugate of the second term. Based on the estimators computed in (\ref{eq:estimators}), we can construct a DS as a linear combination of the square of the norm of these estimators \cite{Astone_2010}, as follows\footnote{We clarify that the expression in \cite{D’Onofrio_2025} is not used. Instead, for simplicity, we adopt the formulation from \cite{Astone_2010}, as this is not a multi-detector search.}:
\begin{eqnarray}\label{eq:DS-def}
\mathcal{S}=|\mathbf{A}^+|^4|\hat{h}^+|^2+|\mathbf{A}^\times|^4|\hat{h}^\times|^2.
\end{eqnarray}
Since in a noiseless scenario the only non-zero Fourier components of a signal are at frequencies $\{f_{\rm GW},f_{\rm GW}\pm f_{\rm sd}, f_{\rm GW}\pm 2f_{\rm sd}\}$, the scalar product between the source templates and the data five-vectors will be non-zero at (up to) \footnote{Depending on the SNR, a correctly demodulated signal may result into less than nine peaks.} nine different frequencies centred at $f_{\rm GW}$ \cite{Singhal_2019}, i.e. at  $f_{\rm GW}-mf_{\rm sd}$ with $m\in [-4,4]$.
Hence, we observe that the five-peaks in \fref{fig:5-peaks} are transformed into a maximum of nine DS-peaks. 
This results from the cross-correlation between a five-peak template and a five-peak signal vector.  

Assuming that the noise is Gaussian and white, with a mean of zero and a variance of $\sigma^2$, from the definition of noise five-vector in \eref{eq:noise-5-v}, it follows that every component of the noise five-vector is also distributed according to a Gaussian with a mean of zero and a variance of $\sigma^2\cdot T_{\rm obs}$, with $T_{\rm obs}$ the total effective observation time. Hence, the probability density $f(\mathcal{S})$ in pure Gaussian noise is \cite{PhysRevD.89.062008}
\begin{eqnarray}\label{eq:s-prob-dens}
f(\mathcal{S}) = \frac{\exp(-\frac{\mathcal{S}}{\sigma^2 T_{\rm obs}|\mathbf{A}^\times|^2})-\exp(-\frac{\mathcal{S}}{\sigma^2 T_{\rm obs}|\mathbf{A}^+|^2})}{\sigma^2 T_{\rm obs}\left(|\mathbf{A}^\times|^2-|\mathbf{A}^+|^2\right)}.
\end{eqnarray}
For known sky location, $(\mathbf{A}^+, \mathbf{A}^\times)$ are source templates that are applied as a match-filter for detecting the signal. Consequently, the computation of the histogram of $\mathcal{S}$ (such that the area under the curve is unity), follows the function in \eref{eq:s-prob-dens}. By integrating \eref{eq:s-prob-dens}, we obtain the probability to have a DS value above a given threshold $\mathcal{S}^\ast$, i.e.:
\begin{eqnarray}\label{eq:prob-s}
\eqalign{
P(\mathcal{S}>\mathcal{S}^\ast) &=\int_{\mathcal{S^\ast}}^\infty f(\mathcal{S}){\rm d} \mathcal{S} =\cr
&=\frac{|\mathbf{A}^\times|^2\exp(-\frac{\mathcal{S}^\ast}{\sigma^2 T_{\rm obs}|\mathbf{A}^\times|^2})-|\mathbf{A}^+|^2\exp(-\frac{\mathcal{S}^\ast}{\sigma^2 T_{\rm obs}|\mathbf{A}^+|^2})}{|\mathbf{A}^\times|^2-|\mathbf{A}^+|^2}.}
\end{eqnarray}

\subsection{Resampling pipeline}
\begin{figure}[H]
    \begin{indented}
    \lineup
    \item[]
    \includegraphics[width=1\linewidth]{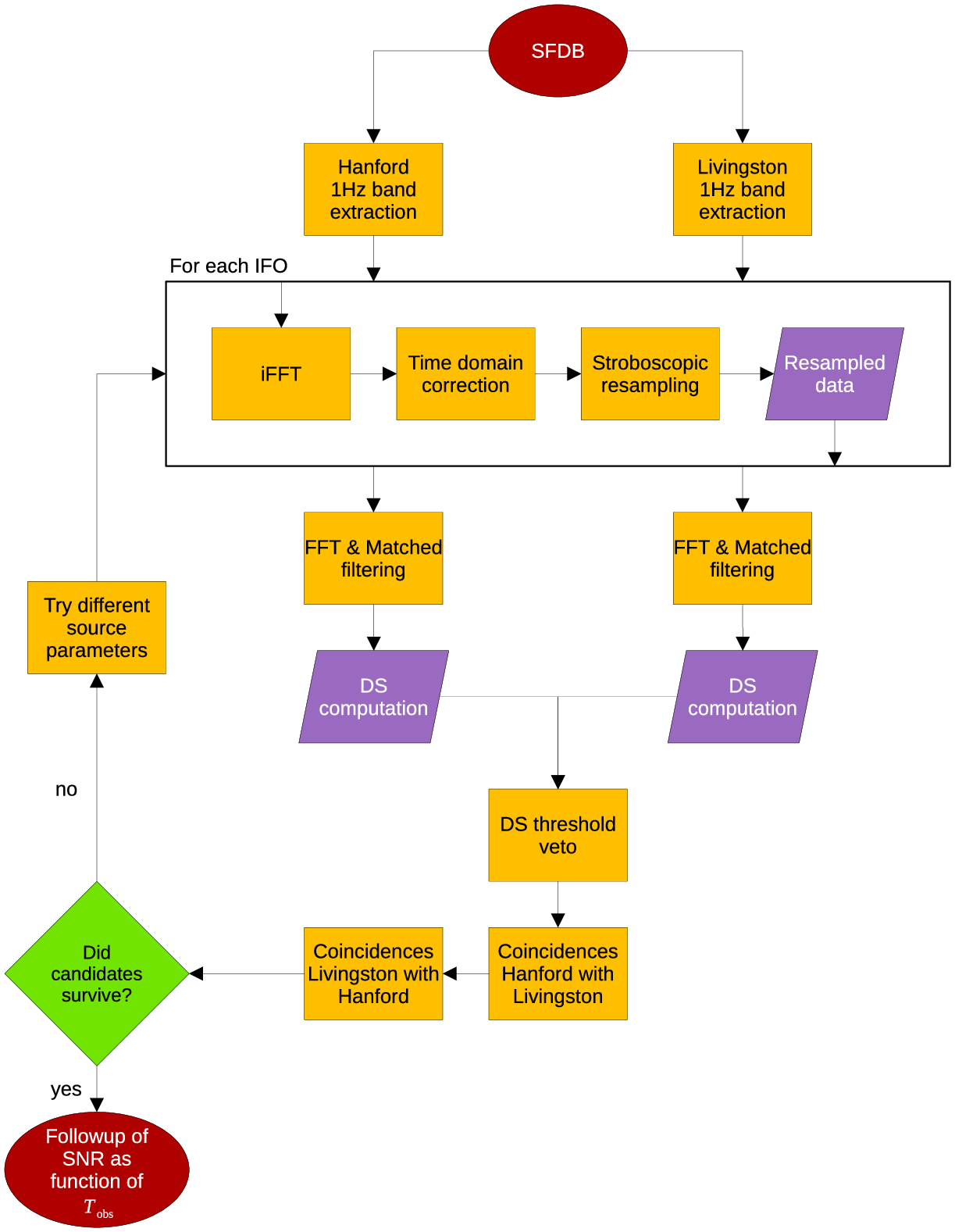}
    \end{indented}
    \caption[Workflow of the binary resampling algorithm to detect CWs from a binary NS]{Workflow of the binary resampling algorithm to detect CWs from a binary NS.}
    \label{fig:workflow}
\end{figure}

In \fref{fig:workflow} we have sketched the resampling pipeline flowchart. It commences from the creation of a Short Fast Fourier Transform (FFT) Database from detector calibrated data, the so-called SFDB \cite{SFrasca_2004, Astone_2005}. From this data, we extract a 1 Hz wide band, which is properly resampled in the time domain. Then, the DS is computed, and coincidences among detectors are performed to filter out false CW candidates (as outlined in \sref{subsec:coincidences}). If no candidate survives the coincidence veto, the resampling procedure can be repeated by adjusting the five Keplerian parameters of the target source used to define the new time variable in \eref{eq:new-time-var}. This step is particularly important when the source ephemerides are not known with sufficient precision, as in the case of Scorpius X-1. In our approach, we varied these parameters by incorporating the maximum allowable offsets needed to detect CWs, as determined in a previous study \cite{Singhal_2019}.
Conversely, if a candidate survives the veto procedure, it necessitates a follow-up analysis to thoroughly assess the candidate and verify its significance. For example, a simple follow-up method involves studying the behavior of the SNR as a function of $T_{\rm obs}$. In the case of a genuine CW signal, the SNR is expected to increase, at best in Gaussian noise, proportionally to the square root of $T_{\rm obs}$ \cite{Riles2023}. A detailed explanation of the veto procedure, first described in \cite{Singhal_2019}, follows.
\subsection{DS veto procedure}
For a given false alarm rate (FAR), we can use \eref{eq:prob-s} to fix a threshold on the DS such that
\begin{eqnarray}\label{eq:NFAR}
P(\mathcal{S}>\mathcal{S}^\ast) \leq \frac{\rm FAR}{\rm \#\,of\,independent\,trials}\equiv {\rm NFAR},
\end{eqnarray}
where the ``\# of independent trials" refers to the number of points in the search parameter space and is equal, in this specific case, to the number of frequency bins times the number of sets of binary search parameters. In this way we account for the \textit{look-elsewhere} effect \cite{Gross2010} (a statistical phenomenon where an apparently significant observation may arise by chance due to the search parameter space size).

A candidate having a DS value higher than the threshold is less likely to be caused by Gaussian noise and needs to be deeply followed up. In the following, this will be called the DS veto. However, the presence of instrumental lines and data gaps cause deviations from this expected behavior. To mitigate their impact on data analysis, these lines are documented and often made publicly available, along with instructions on how to calculate their harmonics. The website~\cite{IL} provides information on the known instrumental lines for the third observation run of the LIGO-Virgo-KAGRA collaboration. These deviations are more prominent towards the tail of the distribution when $\mathcal{S}$ is large and are taken into account in the computation of the threshold, as detailed in \ref{app:ds-thre}. Rather than using $\mathcal{S}$, we follow \cite{Singhal_2019} and compute $\mathcal{S}'$:
\begin{eqnarray}\label{eq:DS-new}
\mathcal{S}'=\frac{\mathcal{S-\mu(\mathcal{S})}}{\sigma(\mathcal{S})}\,,
\end{eqnarray}
which is a linear function of $\mathcal{S}$; $\mu(\mathcal{S})$ and $\sigma(\mathcal{S})$ are the mean and standard deviation of $\mathcal{S}$ across a given 1 Hz band, respectively. Finally, in the real O3 search, we set as a valid threshold a value of ${\mathcal{S}'}^\ast= 18$ (for further details see \ref{app:ds-thre}).

\subsection{Coincidence-based veto procedure}
\label{subsec:coincidences}
To discard candidates not caused by real CWs, we employ a coincident veto criterion described in \cite{Singhal_2019}. Given a candidate at a frequency $f_{\rm cand}$ in a particular interferometer, the three coincidence veto consists of two main steps: 
\begin{enumerate}
    \item \textbf{Internal veto:} compute the coincidence with candidates found within the same detector. A candidate is discarded if there are fewer than $n_{\rm coin}^{\rm int}$ peaks above threshold in the same detector at the frequencies $f_{\rm cand}+mf_{\rm sd},\;m\in[-4,4]$; $n_{\rm coin}^{\rm int}$ is an integer varying  between 1 and 9, but in the O3 search we have set it to three, as in the previous study \cite{Singhal_2019}.
    \item \textbf{Inter-detector veto:} compute the coincidences with candidates identified in the other detector. In the context of simultaneous observations of GW data by multiple detectors with the same $T_{\rm obs}$ and frequency resolution, it is expected that signal peaks will consistently align at the same (up to) nine DS-peak frequencies across all detectors. To distinguish true signals from instrumental noise, a candidate surviving the internal veto is rejected if fewer than $n_{\rm coin}$ peaks surpass the threshold at the frequencies $f_{\rm cand} + m f_{\rm sd}$, where $m \in [-4, 4]$, in the corresponding detector. Here, $n_{\rm coin}$ is an integer ranging from 1 to 9. Candidates failing to meet this criterion are likely to be artefacts. Also in this case, we have set it equal to three. 
\end{enumerate}
To be conservative, however, we use a certain tolerance that considers an extra frequency bin $\Delta f=1/T_{\rm obs}$ on either side of each of the expected nine DS-peaks frequencies, resulting, for a given candidate frequency $f_{\rm cand}$, in a wider nine-peak coincidence interval defined as
\[
\bigcup_{m\in[-4,4]}\left[f_{\rm cand}+mf_{\rm sd}-\Delta f,\; f_{\rm cand}+mf_{\rm sd}+ \Delta f\right].
\]

%% file: 4_Results.tex
\section{Results}\label{results}
The uncertainties in the electromagnetically measured binary period and time of ascending node for Scorpius X-1, as reported in \tref{tab:ScoX1-ephe}, exceed the maximum allowable offsets (in binary parameters) required to detect a genuine CW signal, as shown in ~\cite{Singhal_2019} (see \tref{tab:offset}).
Hence, we used the Scorpius X-1 ephemerides from~\tref{tab:ScoX1-ephe} and offset parameters from \tref{tab:offset} to perform, for the first time, three real narrowband resampling searches for Scorpius X-1 like sources. These searches spanned the entire frequency range of $[10-1000]$ Hz and the full O3 dataset. These 3 searches are enough to cover the full range between them, without additional templates being placed, since the method is robust enough to any combination of offsets within. Unlike \cite{Abbott_2022_scox1}, we did not exclude frequencies affected by narrowband instrumental disturbances~\cite{IL} in our analysis, relying on our robust two-step veto procedure (\sref{resa}) to eliminate false candidates. This approach was validated using the two binary hardware injections~\cite{HI} in the O3 dataset (\ref{appendix:test}). As a result,  we report all candidates, including those within the frequency ranges of O3 known instrumental lines~\cite{IL}.

\begin{table}[h!]
    \caption[90\% accuracy set]{Set of parameter offsets we used for the binary demodulation step~\cite{Singhal_2019}.}
    \label{tab:offset}
    \begin{indented}
\lineup
\item[]
\begin{tabular}{@{}*{5}{l}}
    \br
         $\Delta P$ [ms] & $\Delta a_p$ [mls] & $\Delta e$ & $\Delta \omega$ [$\circ$] & $\Delta t_p$ (s)  \cr
         \mr
         24 & 14 & $10^{-4}$ & 0.01 & 0.1\cr
    \br
    \end{tabular}
    \end{indented}
\end{table}

The three real O3 searches we considered are characterized by null offset (i.e., assuming that the Scorpius X-1 ephemerides are really exact) $\{P,a_p,e,\omega,t_p\}$, a “$+$ offset” $\{P+ \Delta P,a_p+ \Delta a_p,e + \Delta e,\omega+ \Delta \omega,t_p + \Delta t_p\}$ and a “$-$ offset” $\{P- \Delta P,a_p- \Delta a_p,e - \Delta e,\omega- \Delta \omega,t_p - \Delta t_p\}$, respectively. The parameter space of the three sets is shown in \tref{tab:ScoX1}.

\begin{table}[h!]
\caption[O3 narrowband search parameter space]{\label{tab:ScoX1}O3 narrowband search parameter space. We note that $a_p$ might vary from $1.4$ s and $3.25$ ls, but we have arbitrarily chosen $a_p = 1.8$ ls and $\omega = 45.1^\circ$, as done in \cite{Singhal_2019}.}
\begin{indented}
\lineup
\item[]\begin{tabular}{@{}*{6}{l}}

\br   

&&\centre{3}{Values}&\\
\ns
&&\crule{3}&\\
         Parameter & & $-$ offset & null offset & $+$ offset & Units  \cr
         \mr
         frequency & $f_0$ & $[10,1000]$ & $[10,1000]$ & $[10,1000]$ & Hz \cr
         projected semi-major axis & $a_p$ & 1786 & 1800 & 1814 & mls  \cr
         time of periapsis passage & $t_{\rm p}$ & 1078161997.7 & 1078161997.8 & 1078161997.9 & GPS s  \cr
        orbital period & $P$ & 68023.896 & 68023.920 & 68023.944 & s \cr
        eccentricity & $e$ & $ 0.0131$ & $ 0.0132$ & $ 0.0133$ & \cr
        argument of periapsis & $\omega$ & 45 & 45.1 & 45.2& degrees \cr

\br
    \end{tabular}
\end{indented}
\end{table}

\begin{figure}[h!]
    \begin{indented}
    \lineup
    \item[]
    \includegraphics[width=1\linewidth]{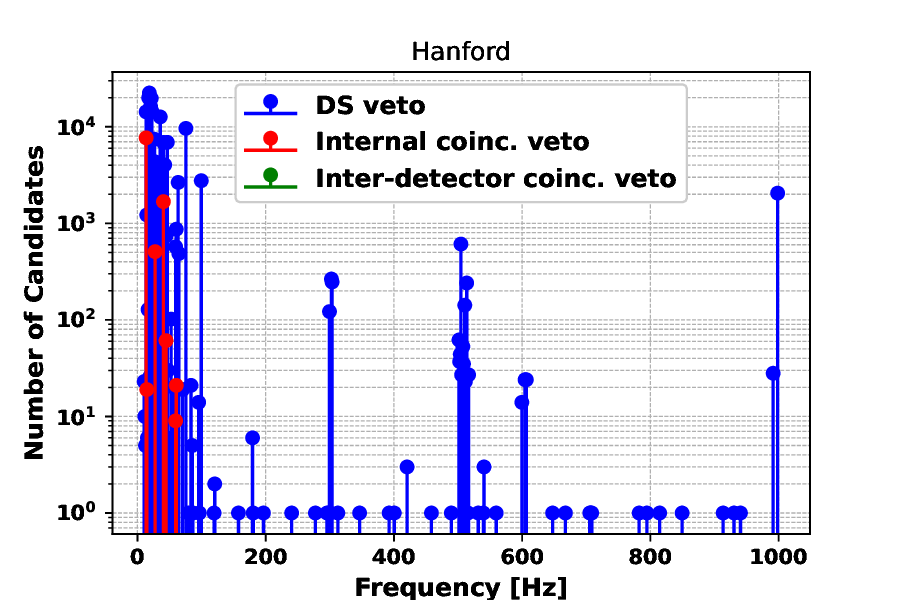}
    \includegraphics[width=1\linewidth]{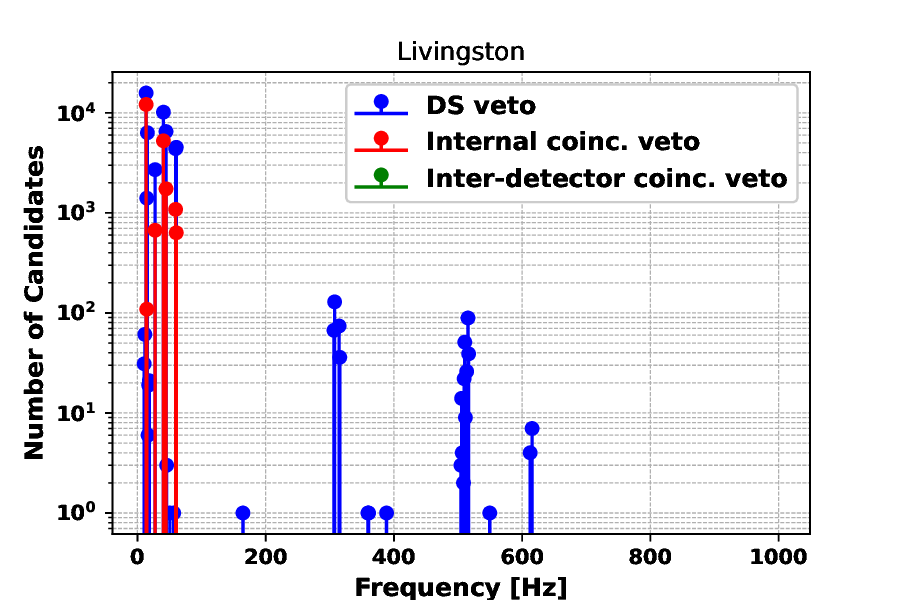}
    \end{indented}
    \caption{Number of candidates surviving the $\mathcal{S'}^\ast=18$ veto for all 1 Hz wide frequency bands in the interval $[10,1000]$ Hz for the real O3 "null offset" search for CWs from Scorpius X-1 in Hanford ({\it top}) and Livingston ({\it bottom}).  
    }
    \label{fig:cand-DS}
\end{figure}

As previously mentioned, we neglect spindown terms. According to the results derived in \sref{resa}, we employ a threshold of $\mathcal{S'}^\ast=18$ to select candidates in every 1~Hz frequency band. Out of approximately $3.06\times 10^{10}$ potential candidates, we identify a total of 184\,726 candidates at Hanford and 52\,527 at Livingston, respectively, when we correct for Doppler orbital modulation with a null offset. The distribution of these candidates among all $1$ Hz frequency bands is shown in \fref{fig:cand-DS}. After applying the three-coincidence veto to this set of surviving candidates, we find that no candidates remain. Similar values are obtained in the '$+$ offset' and '$-$ offset' searches, which are detailed in \tref{tab:surv-cand-H} for Hanford and in \tref{tab:surv-cand-L} for Livingston, respectively..

\begin{table}[H]
\caption{\label{tab:surv-cand-H}Number of candidates in the Hanford detector that survive various veto stages for every search. The $\mathcal{S}'^\ast$ value for the DS veto is 18.}
\begin{indented}
\lineup
\item[]
\begin{tabular}{@{}*{5}{l}}
\br
&&&\centre{2}{Coincidence veto}\\
\ns
&&&\crule{2}\\
Search&Initial candidates&DS veto&Internal&Inter-detector\\
\mr
$-$ offset & $3.05692596 \times 10^{10}$ & 181\,866 & 29\,935 & 0 \cr
null offset & $3.05692596 \times 10^{10}$ & 184\,726  & 9\,979 & 0 \cr
$+$ offset &$3.05692596 \times 10^{10}$ & 182\,946 & 9\,756 & 0 \cr
\br
\end{tabular}
\end{indented}
\end{table}

\begin{table}[H]
\caption{\label{tab:surv-cand-L}Number of candidates in the Livingston detector that survive various veto stages for every search. The $\mathcal{S}'^\ast$ value for the DS veto is 18.
} 
\begin{indented}
\lineup
\item[]
\begin{tabular}{@{}*{5}{l}}
\br
&&&\centre{2}{Coincidence veto}\\
\ns
&&&\crule{2}\\
Search&Initial candidates&DS veto&Internal&Inter-detector\\
\mr
$-$ offset & $3.05692596 \times 10^{10}$ & 50\,479 & 18\,892 & 0 \cr
null offset & $3.05692596 \times 10^{10}$ & 52\,527  & 21\,620 & 0 \cr
$+$ offset &$3.05692596 \times 10^{10}$ & 51\,257 & 20\,753 & 0 \cr
\br
\end{tabular}
\end{indented}
\end{table}

It is important to note that, had any candidates survived the veto chain, we would have performed a thorough follow-up to investigate their origin. This would include using~\cite{universe7070218} in directed mode and passing the information to other pipelines~\cite{Abbott_2022_scox1, PhysRevD.106.062002} for independent verification or further analysis. 

%% file: 5_sensitivity.tex
\section{Sensitivity estimation} 
\label{sec:sensitivity}

We begin by setting upper limits in all 1~Hz frequency bands listed in the first rows of \tref{tab:MC}, performing Monte-Carlo signal-injection studies using the same O3 search and veto-chain schemes. The goal is to find the $h_0$ value such that (at least) 95\% of the signal injections would be recovered with this amplitude value and that are more significant than the most significant candidate identified in the real search in that band. In principle, this should be done by demodulating the injected signals using both the exact binary parameters from~\tref{tab:ScoX1} and the offset parameters from~\tref{tab:offset}. However, for simplicity and to reduce computational costs, we consider the first case and proceed as follows:
\begin{itemize}
    \item In every 1 Hz frequency band in the first rows of \tref{tab:MC}, starting from $H_0= 1\times 10^{-26}$ for the frequency ranges $\{[90,91],\,[100,101],\, [125,126]\}$ Hz and $H_0= 5\times 10^{-26}$ for the other ones, we perform 4600 software injections in the whole O3 data, with source parameters $\{[P, P\pm \Delta P],[a_p, a_p\pm \Delta a_p],[e,e \pm \Delta e],[\omega,\omega\pm \Delta \omega],[t_p,t_p \pm \Delta t_p]\}$, where $\{P, a_{\rm p}, e, \omega, t_{\rm p}\}$ are the Scorpius X-1 ephemeridies of \tref{tab:ScoX1} and $\{\Delta P, \Delta a_{\rm p}, \Delta e, \Delta \omega, \Delta t_{\rm p}\}$ are uniformly drawn from the ranges shown in~\tref{tab:MC}. Frequency values were uniformly  drawn from a subinterval of $0.3$ Hz centered around the midpoint of each 1 Hz analyzed\ band \footnote{This approach is adopted to avoid border effects, which could complicate the detection of signals at the boundary between adjacent bands. }. We also extract $\psi$ and $\cos\iota_{\rm rot}$ from uniform distributions using the range shown in \tref{tab:MC}.
    \item We compute the DS value for every injection data series and select the most significant candidate in the whole 1 Hz band \footnote{The coincidence veto is not considered in computing the upper limits, as no candidates survived the real search.}.
    \item We consider an injected signal as \textit{recovered} if we find a candidate with a value of $\mathcal{S}'$ larger than the largest $\mathcal{S}'$ value found in the real search, and in the same frequency band. 
    \item The above condition must be met at least 95 times (out of 100) for a given amplitude $H_0$ to achieve a 95\% confidence level. If this is the case, we can determine the corresponding upper limit. Otherwise, we repeat the procedure by increasing $H_0$ in steps of $2\times 10^{-26}$.
\end{itemize}
\begin{table}[H]
\caption{Source parameter ranges used to establish upper limits.}
    \label{tab:MC}
\begin{indented}
\lineup
\item[]
\begin{tabular}{@{}*{3}{l}}
\br  
         Parameter & Units & Range \cr
         \mr
         $\Delta f_0$ & Hz & [90,91], [100,101] [125,126], [222,223],\cr
         & & [304, 305], [400,401], [500,501] \cr
         $\Delta a_{\rm p}$ & mls &  [0,14] \cr
         $\Delta t_{\rm p}$ & GPS s & [0, 0.1] \cr
         $\Delta P$ & ms & [0, 24] \cr
         $\Delta e$ & & [0, $10^{-4}$]\cr
         $\Delta \omega$& degree & [0,0.01]\cr
         $\psi$ & rad & [0,$2\pi$]\cr
         $\cos\iota_{\rm rot}$ & & [-1,1]\cr
\br
    \end{tabular}
\end{indented}
\end{table}
To identify the corresponding value of $H_0$ for a given 1 Hz frequency range, we select the two closest values of injected $H_0$, such that one has a recovery rate ($p$) below 95\%, and the other is above. Let these points be $(H_1, p_1)$ and $(H_2, p_2)$, where $p_1<0.95\leq p_2$. We define the slope of the recovery curve as
\begin{eqnarray}
b_{\rm slope} = \frac{H_2-H_1}{p_2-p_1}.
\end{eqnarray}
Using this slope, we perform a linear interpolation to estimate $H_0$ for $p=0.95$, i.e:
\begin{eqnarray}
H_0 = H_1+b_{\rm slope}\cdot(0.95-p_1).
\end{eqnarray}
Finally, we compute the classical strain amplitude $h_0$ via 
\begin{eqnarray}
h_0 \approx 1.31\cdot H_0.
\end{eqnarray}
The factor 1.31 comes from averaging~\eref{eq:Ho-h0} over a flat $\cos\iota_{\rm rot}$ distribution. 

\begin{figure}[H]
    \begin{indented}
    \lineup
    \item[]
    \includegraphics[width=1\linewidth]{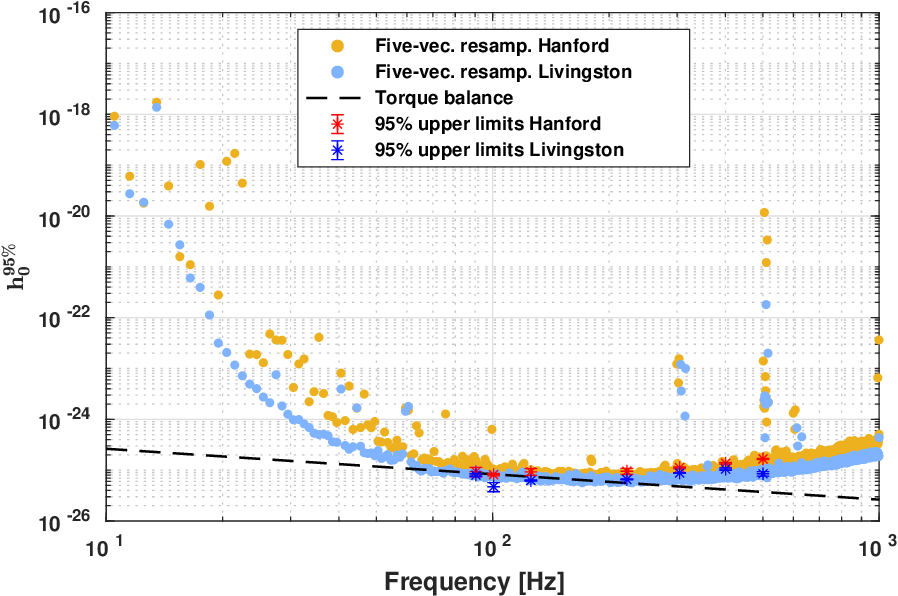}
    \includegraphics[width=1\linewidth]{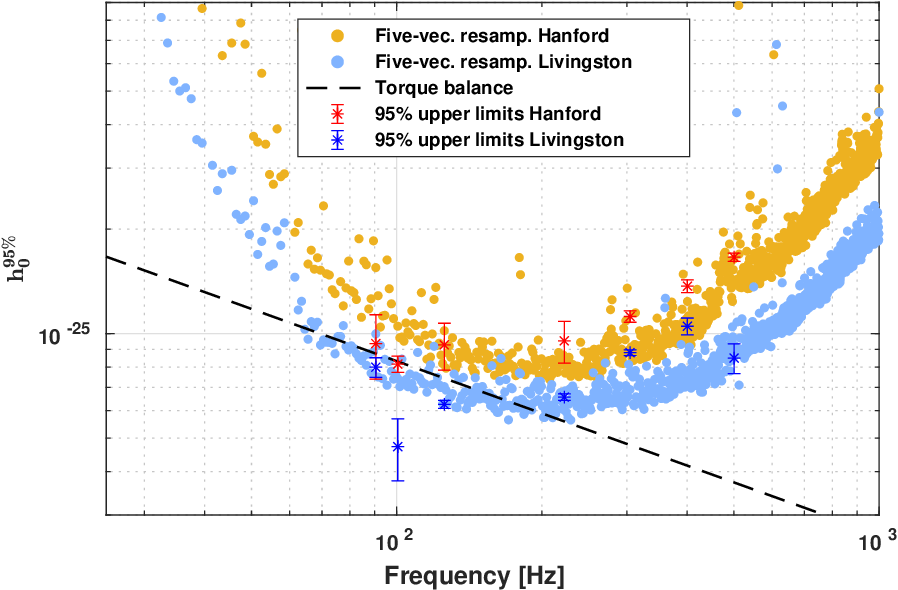}
    \end{indented}
    \caption[Sensitivity estimation]{\textit{Top}: Sensitivity estimation as a function of frequency for the Hanford (orange points) and Livingston (blue points) detectors. These estimates are extrapolated from the computed upper limits (marked with asterisks) with the maximum DS values found in every analyzed 1 Hz-wide band~\cite{PhysRevX.11.021053}. The Upper limits are computed in selected frequency bands and orbital parameter ranges. The vertical error bars indicate the statistical uncertainty on $h_0$, computed as $\Delta h_0 = \sigma_{95\%}\cdot b_{\rm slope}$, where $\sigma_p$ is the binomial standard deviation for a recovery percentage of 95\% using 100 injection: $\sigma_{95\%}=\sqrt{0.95(1-0.95)/100}\sim0.022$. We note that the theoretical strain at the torque-balance level (light green dashed line) is obtained assuming $M_{{\rm NS}}=1.4 M_{\odot}$ and $R_{{\rm NS}}\approx 10\, { \rm km}$. \textit{Bottom:} Zommed section of the above plot in the frequency range $[25, 1000]$ Hz. 
    }
    \label{fig:Sens}
\end{figure}

Upper limits are computed using the CNAF HPC cluster \cite{CNAF}. To extend the analysis to other frequency bands, we apply the interpolation/extrapolation method used in \cite{Singhal_2019}. However, given that our search spans the entire frequency range from 10 Hz to 1000 Hz, we instead utilized the maximum DS values found in every analyzed 1 Hz-wide band, rather than interpolating the upper limits using the interferometer noise sensitivity curve.
The resulting sensitivity estimation versus frequency results are shown in~\fref{fig:Sens}, where it is also shown the theoretical strain (at the torque-balance level) assuming $M_{{\rm NS}}=1.4 M_{\odot}$ and $R_{{\rm NS}}\approx 10 { \rm km}$ \cite{Singhal_2019}
\begin{eqnarray}\label{eq:expc-strain}
h_0 \approx 3.5 \times 10^{-26}\sqrt{\frac{300 {\rm Hz}}{f_{{\rm rot}}}}\,.
\end{eqnarray}
The ``torque balance" curve is included solely to facilitate comparison with existing literature.
As discussed in~\cite{Abbott_2022_scox1}, deriving firm conclusions about the NS equation of state or magnetic field strength from our upper limits is challenging without assuming a specific accretion model. We note that the most stringent sensitivity estimation value is at $f \approx 229.5$ Hz and is equal to $h_0 \approx 5.64\times 10^{-26}$, which is 1.73 times better than the upper limit of the zero-eccentricity cross-correlation search at the same frequency~\cite{Abbott_2022_scox1} and 2.87 times better than the hidden Markov model Viterbi pipeline ~\cite{PhysRevD.106.062002}. However, both the Cross-correlation and Viterbi searches explore the full prior range of orbital parameters, assuming circular orbits, whereas this search focused on a restricted region of the elliptical orbital parameter space.

\begin{figure}[H]
    \label{fig:DvsB}
    \begin{indented}
    \lineup
    \item[]
    \includegraphics[width=1\linewidth]{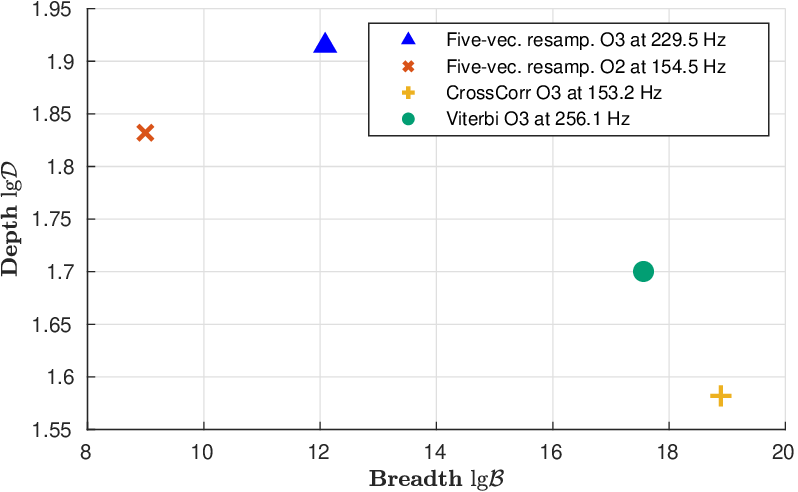}
    \end{indented}
    \caption[]{Sensitivity depth $\lg\mathcal{D}$ versus parameter-space breadth $\lg\mathcal{B}$ computed following~\cite{WETTE2023102880}. The blue point represents the value obtained in this work, with the sensitivity depth $\mathcal{D}$ taken from the most stringent value $h_0 \approx 5.54\times 10^{-26}$ for the Livingston detector at $f_{\mathcal{D}} = 229.5$ Hz. Also displayed are the results for the model-based Cross-correlation search~\cite{Abbott_2022_scox1} and the hidden Markov model Viterbi search~\cite{PhysRevD.106.062002} on O3 data, and the five-vector Resampling search on O2 data \cite{Singhal_2019}, as reported in~\cite{WETTE2023102880}.}
\end{figure}

To account for the different parameter spaces explored by these pipelines, we compute the sensitivity depth $\mathcal{D}$ and the parameter-space breadth $\mathcal{B}$, as defined in Section 4 of~\cite{WETTE2023102880}. The results obtained are $\mathcal{D} = 66.3$ at $f_{\mathcal{D}} = 206.8824$ Hz and $\lg\mathcal{B} = 12.1$. However, when using Equation (68) from~\cite{WETTE2023102880} to compute the binary orbital parameter space, only three out of the five orbital parameters are considered in the metric, namely $\{a_p, P, t_{\rm asc}\}$. Hence, we expect that the actual value of $\mathcal{B}$ for the search presented in this paper would be higher if the remaining two parameters, $\{e, \omega\}$, were also included. In~\fref{fig:DvsB}, we compare our results with those obtained from the Cross-correlation search in O3, the Viterbi search in O3, and the Resampling search in O2 \cite{WETTE2023102880}.

%% file: 6_conclusions.tex
\section{Conclusions}\label{sec:conclusions}
We applied the generalized five-vector resampling methodology to search for CWs from 
NSs in binary systems, focusing on the LMXB Scorpius-X1, across a range of GW signal frequencies $10\,{\rm Hz} \leq f_0 \leq 1000\,{\rm Hz}$, corresponding to NS spin
frequencies of $5\,{\rm Hz} \leq f_{\rm rot} \leq 500\,{\rm Hz}$. This is the first time such a technique is used to perform a real search 
in the O3 advanced LIGO–Virgo-KAGRA dataset. The uncertainties in the electromagnetically measured binary parameters of Scorpius X-1 are significantly larger than the maximum offsets in these parameters that we can use to detect a true  CW signal without deliberately resorting to a matched-filtering approach, as indicated by a previous study \cite{Singhal_2019}. Hence, we restrict our search to three orbital parameter configurations: one assuming the Scorpius X-1 ephemerides $\{P,a_p, e, \omega, t_p\}$ are exact, and two additional configurations where the parameters are adjusted by their respective maximum offsets, i.e., $\{P\pm \Delta P,a_p\pm \Delta a_p,e \pm \Delta e,\omega\pm \Delta \omega,t_p \pm \Delta t_p\}$, with $\{\Delta P = 24\,{\rm ms}, \Delta a_p = 14\,{\rm mls}, \Delta e = 10^{-4}, \Delta \omega = 0.01^\circ, \Delta t_p = 0.1\,{\rm s}\}$ being the maximum values of binary-parameter offsets.

No genuine CW signals can be claimed from these analyses. Therefore, we proceeded to establish 95\% confidence-
level upper limits in selected frequency bands and orbital parameter ranges, while also evaluating overall sensitivity, with the most stringent value being $h_0 \approx 5.64\times 10^{-26}$ at $f \approx 229.5$ Hz for the search with Livingston data.

%% file: 7_Acknowledgments.tex
\ack
This research has made use of data or software obtained from the Gravitational Wave Open Science Center~\cite{PhysRevD.95.062002}, a service of the LIGO Scientific Collaboration, the Virgo Collaboration, and KAGRA. This material is based upon work supported by NSF's LIGO Laboratory which is a major facility fully funded by the National Science Foundation, as well as the Science and Technology Facilities Council of the United Kingdom, the Max-Planck-Society, and the State of Niedersachsen/Germany for support of the construction of Advanced LIGO and construction and operation of the GEO600 detector. Additional support for Advanced LIGO was provided by the Australian Research Council. Virgo is funded, through the European Gravitational Observatory, by the French Centre National de Recherche Scientifique, the Italian Istituto Nazionale di Fisica Nucleare and the Dutch Nikhef, with contributions by institutions from Belgium, Germany, Greece, Hungary, Ireland, Japan, Monaco, Poland, Portugal, Spain. KAGRA is supported by Ministry of Education, Culture, Sports, Science and Technology, Japan Society for the Promotion of Science in Japan; National Research Foundation and Ministry of Science and ICT in Korea; Academia Sinica and National Science and Technology Council in Taiwan.

We would like to acknowledge CNAF for providing the computational resources.

%% file: A_appendix.tex
\section{DS threshold estimation}\label{app:ds-thre}
A critical aspect of the search was determining the correct DS threshold. Assuming the GW data is dominated by Gaussian noise with standard deviation $\sigma$, the distribution of the DS, $\mathcal{S}$, should follow the probability distribution given in \eref{eq:prob-s}. However, instrumental lines and data gaps introduce deviations from this expected behaviour, which are more pronounced in the tail of the distribution, particularly when $\mathcal{S}$ is large. As noise artifacts can significantly influence the DS threshold calculation, it is essential to compute it accurately to reliably identify significant candidates. 

\begin{figure}[H]
    \centering
    \includegraphics[width=1\linewidth]{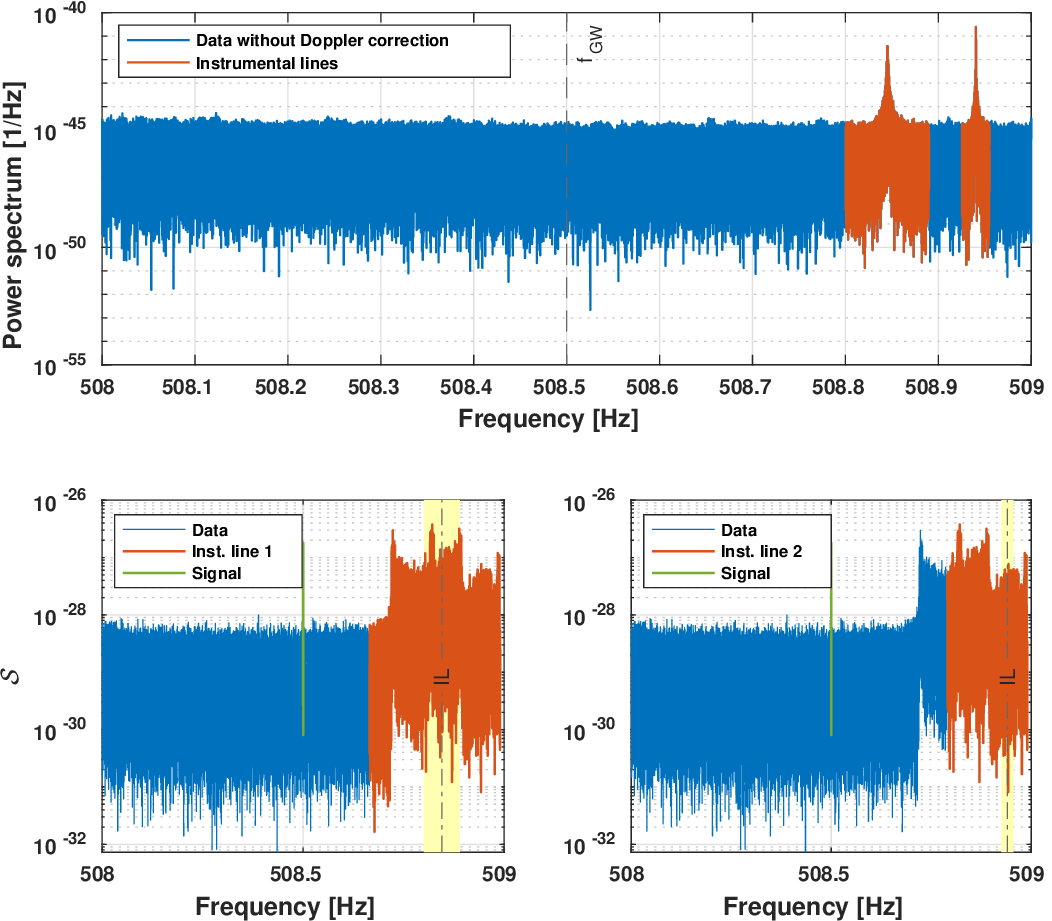} 
    \caption[Broadening of the instrumental lines]{CW signal injected in 10 days of Hanford O3 data, with strain $H_0=1\times 10^{-24}$, $f_{\rm GW}=508.5$ Hz, orbital parameters $P=68023.92 $ s, $a_{\rm p}= 1.8$ ls, $e=0.0132$, $\omega = 45.1^\circ$ and $t_{\rm p}= 1078165521$ GPS. Power spectrum of (both Earth and binary Doppler) modulated data. The two orange instrumental lines at $\sim 508.85$ Hz and $\sim 508.95$ Hz (catalogued both as “Violin mode 1st harmonic” in \cite{IL}) are visible and are marked in orange. \textit{Bottom:} DS after demodulating the data according to the description in \sref{resa}. On the \textit{left} (\textit{right}) we highlight in orange the broadened band of the \textit{first} (\textit{second}) instrumental line. Because of their overlap, the lines are not plotted together. In both cases, the yellow band indicates the original width of the instrumental line for reference.}
    \label{fig:removing_of_IL}
\end{figure}

Using the known instrumental-line frequencies from O3 \cite{IL}, we calculate the Doppler broadening effect, $\Delta M$, induced by binary demodulation corrections and Earth's revolution corrections $\Delta M_{\rm T}$ ($\sim 10^{-4}$), i.e.: \cite{PhysRevD.91.102003}
\begin{eqnarray}
    \Delta M_{\rm tot} = \frac{a_{\rm p}\Omega}{1-e} + \Delta M_{\rm T}.
\end{eqnarray}
This effect shifts the left edge of the broadened region by $\Delta M_{\rm tot} $ to the left and the right edge by $\Delta M_{\rm tot} $ to the right. For example, in \fref{fig:removing_of_IL}, we illustrate this for a 508.5~Hz injected Scorpius X-1-like signal with $H_0=1\times 10^{-24}$ in the band $[508,509]$ Hz. In this case, two instrumental lines have been Doppler broadened due to the demodulation correction, causing them to overlap when computing the DS. 

Rather than removing the entire Doppler-modulated intervals for all instrumental lines, we opted to compute the DS threshold using a dataset where the instrumental lines are set to zero.

Once the instrumental lines are removed, we can compute the normalized histogram of the DS distribution within a given band, which is expected to follow the form given in \eref{eq:s-prob-dens}. However, this analytical function does not provide a good fit for the higher values of $\mathcal{S}$, as the assumption of purely Gaussian noise breaks down at these levels. To account for the non-Gaussian nature of the data, we can approximate the behavior of~\eref{eq:s-prob-dens} at large values of $\mathcal{S}$ with
\begin{eqnarray}\label{eq:s-prob-dens-approx}
    f(\mathcal{S})\sim Ue^{-q\mathcal{S}},
\end{eqnarray}
with $U$ and $q$ two constants that must be obtained fitting the data distribution. 
Additionally, the highly discrete nature of the data makes it challenging to accurately fit the tail of the distribution with \eref{eq:s-prob-dens-approx} or any other reliable representation of the behavior of $f(\mathcal{S})$. As a result, we excluded the largest DS values from the fit, removing the top $15\%$ of the data. Specifically, the fitting procedure was performed on the data interval corresponding to the 45th to 85th percentiles of the DS distribution. This choice was made to avoid the influence of extreme values that could distort the fitting process, as suggested in previous resampling studies.

\begin{figure}[H]
    \begin{indented}
    \lineup
    \item[]
    \includegraphics[width=1\linewidth]{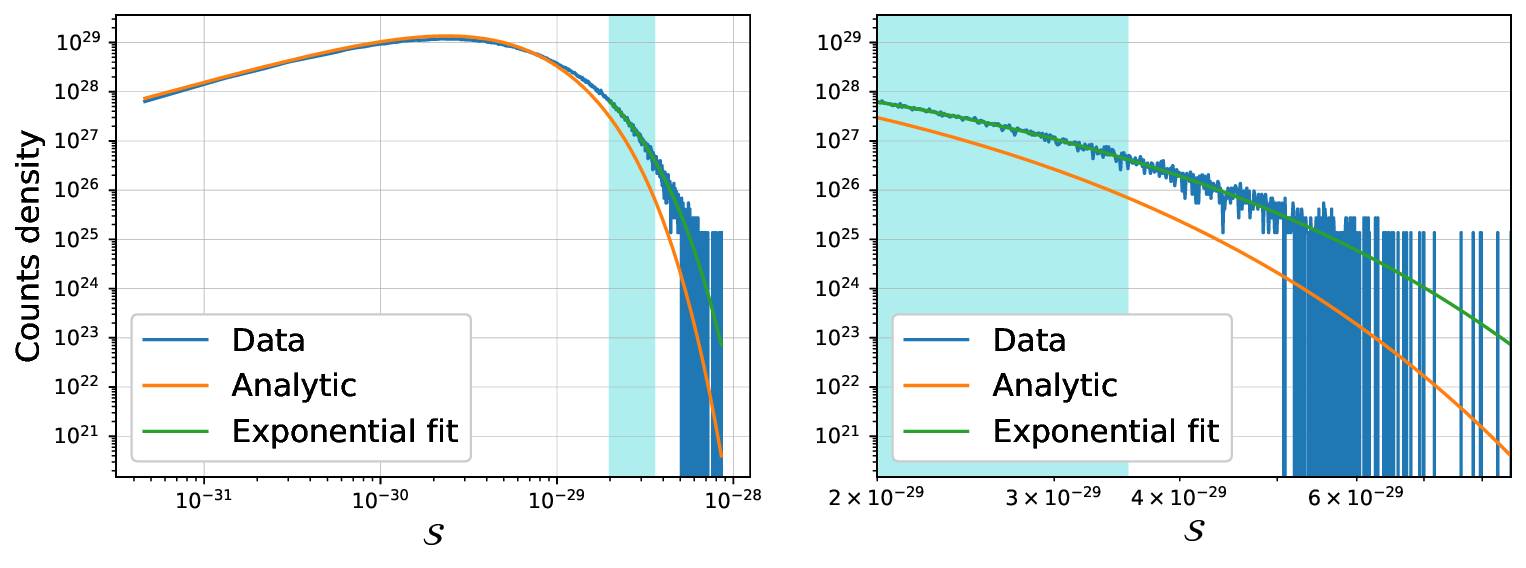}
    \end{indented}
    \caption[]{\textit{Left:} The DS frequency histogram computed for the [508, 509] Hz band, based on the same data from \fref{fig:removing_of_IL}, with instrumental lines removed and no signal injection applied. The analytic function \eref{eq:s-prob-dens} is shown in orange, while the exponential fit using \eref{eq:s-prob-dens-approx} is depicted in green. The region selected for the exponential fit is highlighted in turquoise. \textit{Right:} A zoomed-in view of the left plot around $\mathcal{S} \sim 10^{-29}$.
     }
    \label{fig:analitic-vs-tail}
\end{figure}

As shown in \fref{fig:analitic-vs-tail}, the exponential fit provides a better match to the tail of the distribution in the data presented in \fref{fig:removing_of_IL}. This data was obtained after removing the instrumental lines and without performing any signal injection, highlighting the effectiveness of the fit in capturing the behavior of the tail.
By fitting the function \eref{eq:s-prob-dens-approx} to the data in logarithmic scale, we can determine the constants $U$ and $q$, which are dependent on factors such as the source template, noise characteristics, and potential data artifacts with unknown origins. This allows us to compute the probability of obtaining a DS value larger than a given threshold $\mathcal{S}^\ast$, i.e.: 
\begin{eqnarray}
\label{eq:prop-exp}
P(\mathcal{S}>\mathcal{S}^\ast) =\int_{\mathcal{S^\ast}}^\infty f(\mathcal{S}){\rm d} \mathcal{S}=\frac{Ue^{-q\mathcal{S}^\ast}}{q}.
\end{eqnarray}
However, before using \eref{eq:prop-exp} at a given FAR to compute $\mathcal{S}^\ast$, we need to account for the \textit{look-elsewhere} using \eref{eq:NFAR}. Hence, \eref{eq:s-prob-dens-approx} can be used to compute the threshold $\mathcal{S}^\ast$ as
\begin{eqnarray}\label{eq:thresh}
\mathcal{S^\ast} = -\frac{1}{q}\log(\frac{q}{U}{\rm NFAR}).
\end{eqnarray}

\begin{figure}[H]
    \begin{indented}
    \lineup
    \item[]
    \includegraphics[width=1\linewidth]{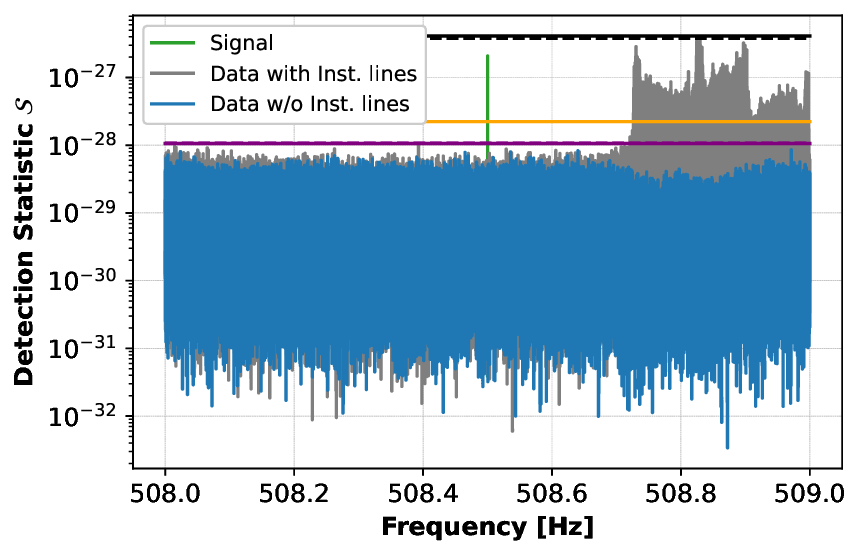}
    \end{indented}
    \caption[Comparison of $\mathcal{S}^\ast$ with and without the removing of the noise]{DS values versus frequency for 10 days of Hanford data, comparing results with (gray lines) and without (blue lines) the removal of instrumental lines, based on the resampled dataset from \fref{fig:removing_of_IL}. The $\mathcal{S}^\ast$ values are shown as horizontal lines for five different cases: a solid black (purple) line at $\mathcal{S}\approx4.09\times 10^{-27}$ ($\mathcal{S}\approx1.07\times 10^{-28}$) for an injected Scorpius X-1-like signal with (without) instrumental lines present; a dashed black (purple) line at $\mathcal{S}\approx3.78\times 10^{-27}$ ($\mathcal{S}\approx1.06\times 10^{-28}$ ) for an injected Scorpius X-1-like signal with (without) instrumental lines removed; and finally, an orange line at $\mathcal{S}\approx2.25\times 10^{28}$ representing the threshold obtained using simulated Gaussian noise data (this dataset is not shown in the plot).
    }
    \label{fig:DS_removing_Il}
\end{figure}

The effect of removing the instrumental lines is illustrated in \fref{fig:DS_removing_Il}, using the same dataset as in \fref{fig:removing_of_IL}.

As shown in \fref{fig:DS_removing_Il}, the removal of instrumental lines has a significant impact on the DS threshold estimation, reducing it by more than an order of magnitude. Specifically, the threshold drops from $\sim 4.09\times 10^{-27}$ (with instrumental lines) to $\sim 1.06\times 10^{-28}$ (after removing the lines), while the effect of the simulated signal is negligible, shifting from $\sim 3.78\times 10^{-27}$ to $\sim 1.07\times 10^{-28}$ in \fref{fig:DS_removing_Il}. For comparison, we also estimated the DS threshold using only Gaussian noise, with an ASD value of $\sim 8.4 \times 10^{-24}~\frac{1}{\sqrt{\mathrm{Hz}}}$ (for the Hanford detector) in the frequency range $[508,509]$ Hz \cite{PhysRevX.11.021053}. This yielded a threshold of $\mathcal{S}^\ast \approx 2.25 \times 10^{-28}$. Therefore, the removal of instrumental lines leads to a DS threshold that is slightly lower than the value obtained from pure Gaussian noise ($\mathcal{S}^\ast \sim 2.25 \times 10^{-28}$). Setting a threshold without first removing the instrumental lines would have prevented the recovery of the injected signal.

\begin{figure}[H]
    \begin{indented}
    \lineup
    \item[]
    \includegraphics[width=1\linewidth]{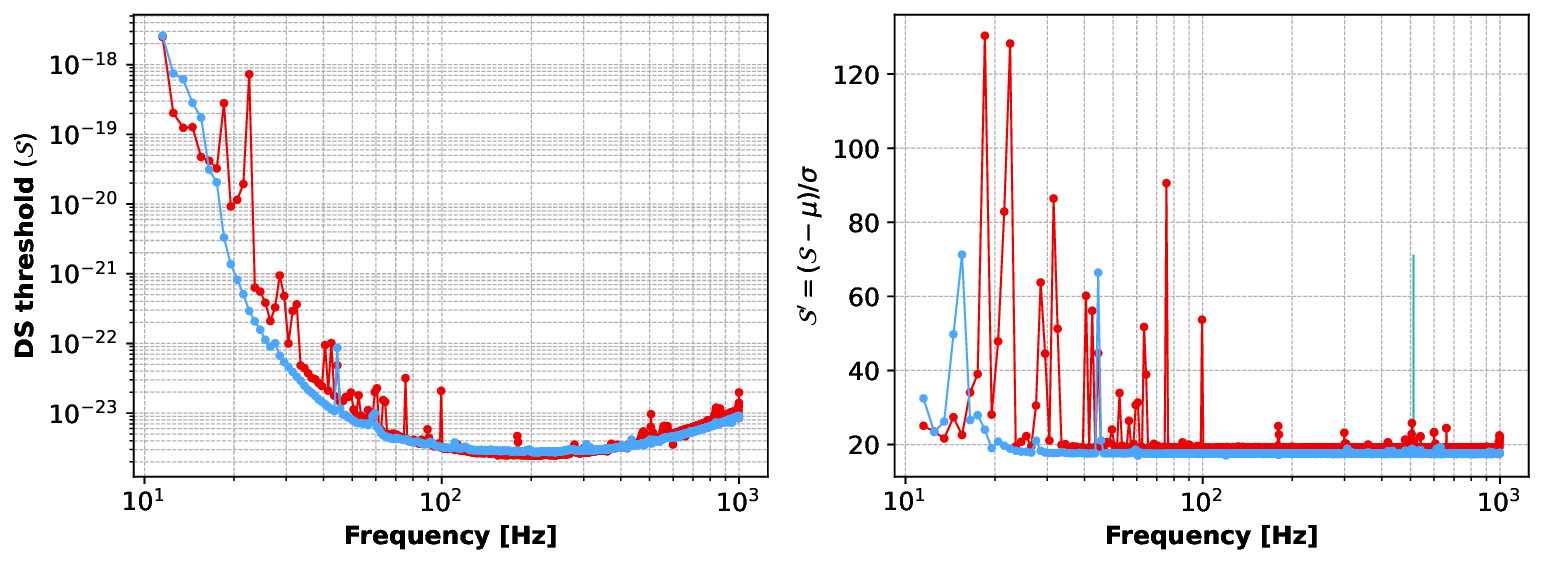}
    \end{indented}
    \caption{DS threshold $\mathcal{S^\ast}$ (\textit{Left plot}) and $\mathcal{S}'^\ast$ (\textit{Right plot}) versus frequency, both for Hanford (red line) and Livingston (blue line) detectors. A data point is obtained for every 1 Hz band, which is then plotted. In the Livingston $[508,515]$ Hz band there are no data to compute the DS threshold as the whole band is contaminated by instrumental lines, which -we remind- have been set to zero only for the DS threshold estimation.}
    \label{fig:DS_thr_mean}
\end{figure}

We applied the procedure described above to the entire frequency band $[10,1000]$ Hz using the real O3 1-year dataset, while setting all known instrumental lines to zero and demodulating the data with the Scorpius X-1 parameters from \tref{tab:ScoX1}. For every 1 Hz band, we observed a wide range of DS threshold values, varying from approximately $\mathcal{O}(10^{-18})$ to $\mathcal{O}(10^{-24})$. Consequently, a rigorous approach would require the use of an adaptive threshold that depends on the specific frequency band being analyzed. Hence, rather than using $\mathcal{S}$, as defined in \cite{PhysRevD.89.062008}, we compute $\mathcal{S}'$ defined in \eref{eq:DS-new}, as introduced in \cite{Singhal_2019}, to better account for the specific characteristics of our analysis.
We use $\mathcal{S'}$ in \eref{eq:DS-new} because the noise-only distribution is normalized in $\mathcal{S'}$ \cite{PhysRevD.90.042002}, which enhances the robustness of the analysis across multiple detectors and varying frequency bands. The values of $\mathcal{S}$ and $\mathcal{S}'$ as a function of the frequency are shown in \fref{fig:DS_thr_mean}. 
As an example, in the $[100,130]$ Hz band, which is relatively clean, the mean value of ${\mathcal{S}'}^\ast$ for the two detectors is
\begin{eqnarray}
    {\rm Hanford:} &\quad \langle{\mathcal{S}'}^\ast\rangle\approx 19.2,\\
    {\rm Livingston:} &\quad  \langle{\mathcal{S}'}^\ast\rangle\approx 17.6.
\end{eqnarray}
Therefore, we choose $\mathcal{S}' = 18$ as a reasonable threshold value for the real O3 search. This value is consistent with the results reported in \cite{Singhal_2019} for a previous LIGO-Virgo-KAGRA run. It is important to emphasize that this check is essential, as the DS threshold, ${\mathcal{S}'}^\ast$, must be recalculated each time a new LIGO-Virgo-KAGRA observing run dataset is analyzed.

%% file: B_appendix.tex
\section{Testing with Hardware injections}
\label{appendix:test}
To validate the implementation of the resampling technique, we use  hardware injections present in O3 data \cite{PhysRevD.95.062002}. These injections refer to simulated GW signals introduced by physically displacing the test masses of detectors, effectively mimicking their response to real GW signals \cite{PhysRevD.95.062002}. During the O3 observing run, 17 hardware injections simulating CW signals from rapidly rotating NSs were introduced in the two LIGO detectors. The specific parameters for each injected pulsar are listed in \cite{HI}. Out of the 17 injected signals, only pulsars 16 and 17 are associated with binary systems, while the others simulate isolated NS signals. The detailed injection parameters are provided in \tref{tab:HI}. Notice that the injected binary pulsars have no spindown terms, although spindown term contributions are, in any case, already well implemented in the resampling pipeline.

\begin{table}[H]
\caption{\label{tab:HI} Parameters of hardware-injected binary pulsars in the LIGO Detectors.} 

\begin{indented}
\lineup
\item[]
\begin{tabular}{@{}*{4}{l}}
\br                              
Parameter & Units& Pulsar 16 & Pulsar 17 \cr 
\mr
$\alpha$ & rad & -0.2729743 & -0.2729743 \cr
$\delta$ & rad & 0.3487022  & 1.9194985 \cr
$f_{\rm GW}$ & Hz & 234.567 & 890.123 \cr
$h_0$ & & $\sim 1.59 \times 10^{-25}$& $\sim 8.39\times 10^{-26}$ \cr
$\psi$ & rad & 4.084 & 4.084 \cr
$\iota_{\rm rot}$ & rad & 0.7598222219515758 & 0.7675866198953613 \cr
$a_p$    &  ls  & 2.35 & 2.35 \cr
$t_{\rm p}$ & GPS s & 1230336018 &  1230379218  \cr
$P$ & s & 83941.2 & 83941.2 \cr
$e$ &  & 0.0 & 0.0 \cr
$\omega$ & rad & 1.23456 & 2.13456 \cr
\br
\end{tabular}
\end{indented}
\end{table}

\begin{figure}[H]
    \begin{indented}
    \lineup
    \item[]
    \includegraphics[width=1\linewidth]{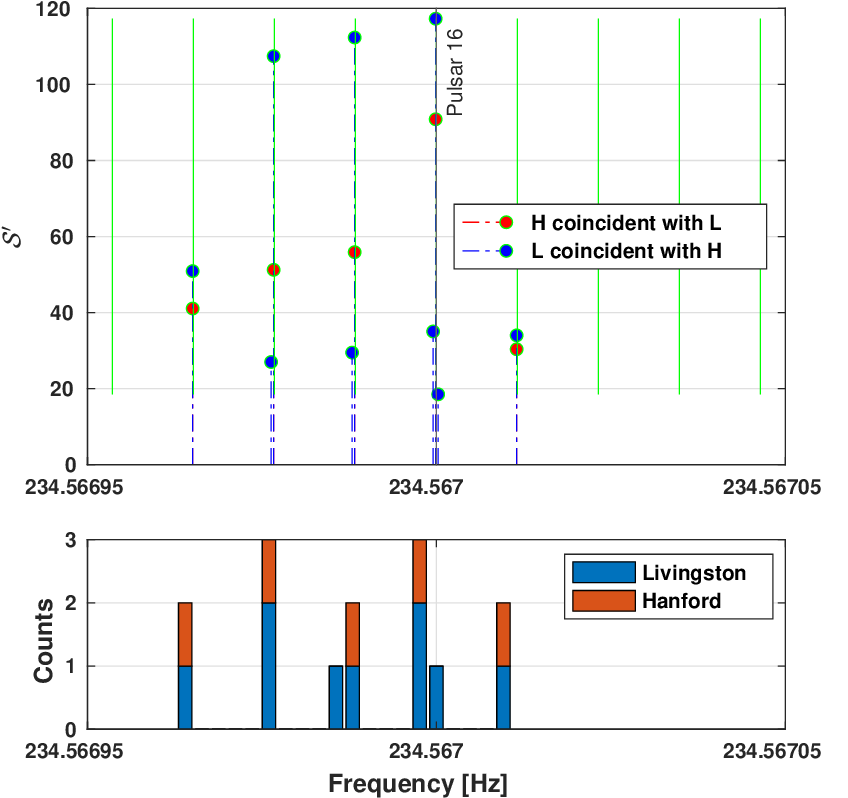}
    \end{indented}
    \caption{\textit{Top}: The $\mathcal{S}'$ values for candidates that survived the coincidence veto with $n_{\rm coin} = 3$ and $\mathcal{S}'^\ast = 18$ are shown after demodulating the data using the pulsar 16 hardware injection parameters. The green vertical lines mark the expected frequencies of the nine peaks, while the vertical black line at $234.567$ Hz indicates the frequency of the hardware injection. \textit{Bottom}: Stacked histogram showing the frequency distribution of the candidates for both Hanford and Livingston detectors. }
    \label{fig:6}
\end{figure}

Using data from the first 31 days of the O3 run, we successfully detected pulsar 16 by identifying one or more candidates that survived to the three coincidence veto, and that are found in five of the nine expected frequency bins (see \fref{fig:6}). However, despite analyzing the full O3 dataset, we did not detect pulsar 17. While we identified three candidates surviving the DS veto in Livingston and one in Hanford, these candidates did not pass the coincidence veto, preventing a confirmed recovery.

%% file: C_appendix.tex
\section{Computational cost}
\label{appendix:Abis}
We detail the computational budget required by the resampling algorithm. The bulk of the computation was performed on the CNAF cluster, using various x86\_64 architecture CPUs, including the AMD EPYC 7351 16-Core Processor. Each detector and 1 Hz frequency band required approximately 14.5 GB of memory (RAM + swap) for processing \cite{CNAF}. The total computational time, $\mathcal{C}_{\rm tot}$, and disk space, $\mathcal{D}_{\rm tot}$, can be estimated as the sum of computational resources required ofr each individual step in the analysis: extracting the 1 Hz bands ($\mathcal{Y}_{\rm 1Hzband}$), performing the time-domain resampling ($\mathcal{Y}_{\rm resamp}$), computing the DS ($\mathcal{Y}_{\rm DS}$), and applying the coincidence veto ($\mathcal{Y}_{\rm coinc}$). Hence, we can express the total resources as
\begin{eqnarray}
 \mathcal{C}_{\rm tot}&=\mathcal{C}_{\rm 1Hzband}+\mathcal{C}_{\rm resamp}+\mathcal{C}_{\rm DS}+\mathcal{C}_{\rm coinc}\label{eq:ctot}\\
 \mathcal{D}_{\rm tot}&=\mathcal{D}_{\rm 1Hzband}+\mathcal{D}_{\rm resamp}+\mathcal{D}_{\rm DS}+\mathcal{D}_{\rm coinc}\label{eq:dtot}
\end{eqnarray}
where $\mathcal{Y}\in \{\mathcal{C},\mathcal{D}\}$. For $T_{\rm obs}=1$ year in the O3 run, for 1~Hz band, the computing cost values we obtained are reported in \tref{tab:tim}. The values of $\mathcal{D}$ are null in the real search and depend on the number of identified candidates. However, they can be estimated to be approximately 1.2 kB per Hz band if any candidates survive the three-coincidence veto in that band.
\begin{table}[H]
    \caption[Time computational cost]{The CPU time and disk space required for the various steps outlined above were estimated based on a real one-year O3 search conducted for each 1 Hz band in a single detector.}
    \label{tab:tim}
\begin{indented}
\lineup
\item[]
\begin{tabular}{@{}*{3}{l}}
\br  
         $\mathcal{Y}$ & $\mathcal{C}$ & $\mathcal{D}$ \cr
         \mr
         $\mathcal{Y}_{\rm 1Hzband}$ & 32.97 m & 351 MB \cr 
         $\mathcal{Y}_{\rm resamp}$ & 6.87 h & 336 MB \cr 
         $\mathcal{Y}_{\rm DS}$ & 3.11 m & 1.5 GB \cr 
         $\mathcal{Y}_{\rm coinc}$ & 20.3 s & // \cr
         \br
    \end{tabular}
    \end{indented}
\end{table}
We note that the worst impact on the time cost comes from the resampling step. As explained in \cite{Singhal_2019}, the computational resource needed to extract 1 Hz band can be expressed as
\begin{eqnarray}
    \mathcal{C}_{\rm 1Hzband} &= c_{\rm 1Hzband} N_{\rm f}N_{\rm det}T_{\rm obs}[{\rm d}]\label{eq:c_sbl},\\
    \mathcal{D}_{\rm 1Hzband} &= d_{\rm 1Hzband} N_{\rm f}N_{\rm det}T_{\rm obs}[{\rm d}]\label{eq:d_sbl},
\end{eqnarray}
with $N_{\rm f}$ the number of 1 Hz frequency bands, $N_{\rm det}$ the number of detectors and $T_{\rm obs}$ the observation time in days.  Hence, using the values of \tref{tab:tim} with $N_{\rm f}=1$, $N_{\rm det}=1$ and $T_{\rm obs}=361$ days, i.e., the number of days between the start and the end of O3, in \eref{eq:c_sbl} and \eref{eq:d_sbl}, we get $c_{\rm 1Hzband}=5.48\,{\rm s}$ and $d_{\rm 1Hzband}=970\,{\rm kB}$. For the steps following the 1 Hz band extraction, the computational resources required (both in terms of computing time and disk space) grow with the number of points $N_{\rm p}$ in the search parameter space. For example, in the research presented in this paper, we covered three set of parameter space, hence $N_p = 3$. Therefore we have:
\begin{eqnarray}
    \mathcal{C}_{\rm resamp} &= c_{\rm resamp} N_{\rm f}N_{\rm det}T_{\rm obs}[{\rm d}]N_{\rm p}\label{eq:c_resamp},\\
    \mathcal{D}_{\rm resamp} &= d_{\rm resamp} N_{\rm f}N_{\rm det}T_{\rm obs}[{\rm d}]N_{\rm p}\label{eq:d_resamp},\\
    \mathcal{C}_{\rm DS} &= c_{\rm DS} N_{\rm f}N_{\rm det}T_{\rm obs}[{\rm d}]N_{\rm p}\label{eq:c_DS},\\
    \mathcal{D}_{\rm DS} &= d_{\rm DS} N_{\rm f}N_{\rm det}T_{\rm obs}[{\rm d}]N_{\rm p}\label{eq:d_DS},\\
    \mathcal{C}_{\rm coinc} &= c_{\rm coinc} N_{\rm f}N_{\rm det}T_{\rm obs}[{\rm d}]N_{\rm p}\label{eq:c_coinc}
\end{eqnarray}
Using again the values in \tref{tab:tim} with $N_{\rm p}=1$ in \eref{eq:c_resamp}-\eref{eq:c_coinc}, we get $c_{\rm resamp}=68.6\,{\rm s}$, $d_{\rm resamp}=930 \,{\rm kB}$, $c_{\rm DS}=0.52\,{\rm s}$, $d_{\rm DS}=4.16 \,{\rm MB}$ and $c_{\rm coinc}=8.61\,{\rm ms}$.

For the search presented in this paper, considering one set of parameters and considering both detectors, i.e., $N_f = 990,\, N_{\rm p}=1 ,\, N_{\rm det}=2,\, T_{\rm obs}=361 {\rm d}$, we get
$\mathcal{C}_{\rm tot} = 148\,195 $ CPU h and $\mathcal{D}_{\rm tot}= 4.33$ TB.